\documentclass[10pt, onecolumn, conference]{IEEEtran}

\usepackage{graphicx,url,tabularx,booktabs}
\usepackage[english]{babel}
\usepackage[utf8]{inputenc} 
\usepackage{xspace}
\usepackage{float}

\newcommand{\toolurl}
{\url{https://github.com/AnonShield/Vulnerability_Extractor}\xspace}

\title{Structured Extraction of Vulnerabilities in OpenVAS and Tenable WAS Reports Using LLMs}

\author{
\IEEEauthorblockN{
Beatriz Machado\IEEEauthorrefmark{1},
Douglas Lautert\IEEEauthorrefmark{1},
Cristhian Kapelinski\IEEEauthorrefmark{1},
Diego Kreutz\IEEEauthorrefmark{1}
}
\IEEEauthorblockA{\IEEEauthorrefmark{1}AI Horizon Labs, Federal University of Pampa (UNIPAMPA) \\ 
\{beatrizmachado, douglaslautert, cristhianavilla\}.aluno@unipampa.edu.br, diegokreutz@unipampa.edu.br}
}

\begin{document} 

\maketitle

\begin{abstract}
    This paper proposes an automated LLM-based method to extract and structure vulnerabilities from OpenVAS and Tenable WAS scanner reports, converting unstructured data into a standardized format for risk management. In an evaluation using a report with 34 vulnerabilities, GPT-4.1 and DeepSeek achieved the highest similarity to the baseline (ROUGE-L greater than 0.7). The method demonstrates feasibility in transforming complex reports into usable datasets, enabling effective prioritization and future anonymization of sensitive data.
\end{abstract}

\begin{IEEEkeywords}
Vulnerability Extraction, OpenVAS, Tenable WAS, Large Language Models, Information Security, Automated Analysis, Chunking Strategies, Report Standardization, Cybersecurity Datasets
\end{IEEEkeywords}

\section{Introduction}
\label{sec:introdução}

Vulnerability scanners such as OpenVAS and Tenable~WAS are widely used to identify flaws in web applications; however, they produce structurally heterogeneous reports, which complicates automated analysis and integration with machine learning models. This inconsistency, combined with the large volume of reported vulnerabilities, intensifies the challenge of prioritization, especially in institutions with limited resources. Data from SGIS\footnote{\url{https://tinyurl.com/RelatorioRNP2017}} indicate that more than 500,000 vulnerabilities remained unaddressed in 2017, a trend that increases with the expansion of the attack surface and the continuous growth of threats, as highlighted by the RNP Security Report\footnote{\url{https://www.rnp.br/publicacao/relatorio-anual-de-seguranca-de-2023}}.

In this context, this work proposes a method based on Large Language Models (LLMs) to extract vulnerabilities from OpenVAS and Tenable~WAS reports. The developed tool, called \textit{Vulnerability Extractor}\footnote{\toolurl}, converts these reports into a structured format to enable subsequent analysis, prioritization, and the construction of datasets usable by machine learning models. The approach aims to standardize and harmonize the extracted information, ensuring consistency across different tools and significantly reducing manual processing overhead. Additionally, the method is designed for future integration with anonymization modules, enabling the creation of secure and shareable datasets among institutions.

\section{Vulnerability Reports}
\label{sec:relatórios}

OpenVAS and Tenable~WAS are among the leading tools for automatic vulnerability detection, but they present structural and semantic differences that directly impact the extraction and analysis of results. Understanding these distinctions is essential for ensuring greater consistency in information extraction and data standardization from both sources.

\begin{table}[H]
\centering
\caption{Fields present in OpenVAS reports.}
\label{tab:openvas_estrutura}
\renewcommand{\arraystretch}{1.3}
\scriptsize
\begin{tabular}{|p{0.19\linewidth}|p{0.7\linewidth}|}
\hline
\textbf{Element} & \textbf{Description} \\ \hline
\textbf{Summary} & Brief summary of the identified issue. \\ \hline
\textbf{Vulnerability Detection Result} & Indicates vulnerable URLs, ports, or services detected by the scanner. \\ \hline
\textbf{Impact} & Explains the potential consequences of exploiting the vulnerability. \\ \hline
\textbf{Solution} & Provides recommendations to mitigate or remediate the issue. \\ \hline
\textbf{Affected Software/OS} & Identifies the affected software, operating system, or component. \\ \hline
\textbf{Vulnerability Insight} & Details the origin and exploitation mechanism of the vulnerability. \\ \hline
\textbf{Vulnerability Detection Method} & Describes the method or script used by the scanner to identify the flaw. \\ \hline
\textbf{Log Method} & Specifies techniques or logging approaches used during detection. \\ \hline
\textbf{References} & Lists CVE identifiers, links, or other relevant references. \\ \hline
\end{tabular}
\normalsize
\end{table}

OpenVAS, as shown in Table~\ref{tab:openvas_estrutura}, prioritizes a high level of technical detail, including fields such as \textit{Vulnerability Insight}, which describe the nature and causes of vulnerabilities \cite{Greenbone_OPENVAS_REPORT}. On the other hand, Tenable~WAS (see Table~\ref{tab:tenable_estrutura}) adopts a more risk-management-oriented approach, incorporating sections such as \textit{Risk Information}, which directly support remediation prioritization \cite{Tenable_WAS_User_Guide}.

Table~\ref{tab:optionsbleed_comparison} illustrates how the same vulnerability may be represented differently across vulnerability scanners, reflecting variations in structure and detail. While OpenVAS tends to use a more granular and technically oriented organization, Tenable~WAS emphasizes consolidated information with a focus on risk management and prioritization. This structural heterogeneity required adaptations in the extraction process to ensure correct interpretation and standardization of data from both tools.

\begin{table}[H]
\centering
\caption{Fields present in Tenable WAS reports.}
\label{tab:tenable_estrutura}
\renewcommand{\arraystretch}{1.2}
\scriptsize
\begin{tabular}{|p{0.19\linewidth}|p{0.7\linewidth}|}
\hline
\textbf{Element} & \textbf{Description} \\ \hline
\textbf{Affected Application} & Information about the affected application, including name, first and last detection dates. \\ \hline
\textbf{Description} & Details of the Tenable plugin responsible for identifying the vulnerability and contextualizing its behavior. \\ \hline
\textbf{Solution} & Recommended mitigation actions, including official fixes when available. \\ \hline
\textbf{See Also} & External links and references that expand the technical description. \\ \hline
\textbf{Vulnerability Properties} & General properties such as severity, exploitability, publication date, and remediation status. \\ \hline
\textbf{Discovery} & Records the first and last detection dates, providing monitoring history. \\ \hline
\textbf{VPR Key Drivers} & Factors used to calculate the \textit{Vulnerability Priority Rating} (VPR), such as impact and exploitation trends. \\ \hline
\textbf{Plugin Details} & Technical information about the detection plugin, including version and dependencies. \\ \hline
\textbf{Risk Information} & Associated risk metrics (CVSSv2, CVSSv3, CVSSv4), attack vectors, and risk modifications such as \textit{Accept} or \textit{Recast}. \\ \hline
\textbf{Reference Information} & External references related to the vulnerability, exploit, patch, or security bulletins. \\ \hline
\end{tabular}
\normalsize
\end{table}

% To handle the diversity of fields and structural differences in reports from both vulnerability scanners, we implemented an explicit mapping in the prompt sent to the language model, associating tool-specific fields with a set of generalized labels. Missing fields were filled with \texttt{NULL} values to prevent the model from generating artificial information and to ensure consistency and fidelity to the original data.

To handle the diversity of fields and structural differences in reports from both vulnerability scanners, we implemented an explicit mapping in the prompt sent to the language model, associating tool-specific fields with a set of generalized labels. Missing fields were filled with \texttt{NULL} values to prevent the model from generating artificial information and to preserve fidelity to the original data. This mapping, documented in the Vulnerability Extractor documentation, ensures consistency across reports and defines the unified schema used to normalize fields from different scanners.

\begin{table}[H]
\centering
\caption{Comparison of OptionsBleed in OpenVAS and Tenable WAS.}
\label{tab:optionsbleed_comparison}
\renewcommand{\arraystretch}{1.2}
\scriptsize
\begin{tabular}{|p{0.2\linewidth}|p{0.33\linewidth}|p{0.33\linewidth}|}
\hline
\textbf{Element} & \textbf{OpenVAS} & \textbf{Tenable WAS} \\ \hline
\textbf{Vulnerability name} & Apache HTTP Server OPTIONS Memory Leak Vulnerability (OptionsBleed) & Apache 2.4.x \textless{} 2.4.28 HTTP Vulnerability (OptionsBleed) \\ \hline
\textbf{CVE} & CVE-2017-9798 & CVE-2017-9798 \\ \hline
\textbf{Description} & Apache HTTP Server allows remote attackers to read data [\dots] & Versions of Apache 2.4.x prior to 2.4.28 are affected by a vulnerability [\dots] \\ \hline
\textbf{Installed version} & 2.2.8 & 2.4.7 \\ \hline
\textbf{Fixed version} & 2.4.28 (or equivalent patch for 2.2.34) & 2.4.28 \\ \hline
\textbf{Impact} & Allows unauthorized reading of memory blocks from the server. & \url{https://httpd.apache.org/security/vulnerabilities_24.html\#2.4.28} \\ \hline
\textbf{Severity (CVSS)} & 5.0 (Medium) & 7.5 (High, CVSSv3) \\ \hline
\textbf{Solution} & Update to Apache HTTP Server 2.4.28 [\dots] & Update to Apache HTTP Server 2.4.28 [\dots] \\ \hline
\textbf{Detection method} & Apache Web Server Detection (OID: 1.3.6.1.4.1.25623.1.0.900498) & Plugin ID 98913 \\ \hline
\textbf{Family / Category} & Web Server Vulnerability & Component Vulnerability \\ \hline
\textbf{References} & \url{http://openwall.com/lists/oss-security/2017/09/18/2 }[\dots] & \url{https://httpd.apache.org/security/vulnerabilities_24.html\#2.4.28} [\dots] \\ \hline
\end{tabular}
\normalsize
\end{table}

\section{Related Work}
\label{sec:relacionados}

As summarized in Table~\ref{tab:kie_llm_recentes}, automated extraction of unstructured data from technical documents has become a relevant advancement in cybersecurity, especially for vulnerability identification and analysis. Recent studies demonstrate that LLMs such as GPT-4, LLaMA, and Claude improve both precision and semantic consistency in various information extraction scenarios. Works such as \cite{FabacherEtAl2025_MedicationNotes} show significant gains in clinical contexts, while \cite{LiEtAl2024_VisualIE} and \cite{ZhongEtAl2024_MultimodalPrompt} explore the potential of LLMs in visual and multimodal documents, highlighting their versatility in integrating text and images. Approaches such as \cite{HuEtAl2025_TransferableKIE}, complemented by analyses from \cite{Chen2025_PromptEngineeringReview}, reinforce the models' ability to generalize extraction tasks, their high adaptability, and efficiency in cross-domain transfer.

\begin{table}[H]
\centering
\caption{Information extraction with LLMs}
\label{tab:kie_llm_recentes}
\renewcommand{\arraystretch}{1.3}
\scriptsize
\begin{tabular}{|p{0.2\linewidth}|p{0.2\linewidth}|p{0.2\linewidth}|p{0.2\linewidth}|}
\hline
\textbf{Reference} & \textbf{Domain / Application} & \textbf{Document Type} & \textbf{Model (LLMs)} \\ \hline

\cite{HuEtAl2025_TransferableKIE} & General / multiple KIE domains & Tables + non-standard layouts & GPT-4, T5-XXL, LLaMA-3 \\ \hline

\cite{FabacherEtAl2025_MedicationNotes} & Clinical / multilingual medical notes & Clinical text (French and English) & GPT-4, LLaMA-3, Mistral-7B \\ \hline

\cite{ZhongEtAl2024_MultimodalPrompt} & Multimodal / image-text & Multimodal AI documents & CLIP, BLIP-2, GPT-4V \\ \hline

\cite{Chen2025_PromptEngineeringReview} & General AI / prompt engineering & Review / survey & GPT-4, Gemini 2, Claude 3, LLaMA-3, Mistral-Large \\ \hline

\cite{YanEtAl2025_DocExtractNet} & Technical / document processing & Technical documents and forms & T5, BART, LLaMA-2, GPT-4 \\ \hline

\textbf{This work} & \textbf{Vulnerabilities / \textit{Chunking}} & \textbf{Vulnerability reports (OpenVAS / Nessus)} & \textbf{GPT-4, GPT-4.1, LLaMA-3, LLaMA-4, DeepSeek} \\ \hline

\end{tabular}
\normalsize
\end{table}

Despite these advances, the current literature focuses mainly on clinical, multimodal, or generic technical documents, with limited emphasis on operational security scanners. This work differs by specifically investigating standardized extraction of vulnerabilities from heterogeneous OpenVAS and Tenable~WAS reports, a domain in which structural and semantic differences directly affect analysis and prioritization. Furthermore, we present explicit field mapping, integration with chunking strategies, and guidelines for generating consistent and anonymizable datasets, aspects still underexplored in the existing literature.

\section{Extraction Pipeline Using LLMs}
\label{sec:pipeline}

The developed tool automates the extraction of vulnerabilities from PDF reports generated by OpenVAS and Tenable~WAS vulnerability scanners, using LLMs to convert unstructured text into structured representations. The main pipeline is organized into modular stages that ensure integrity and consistency of the extracted data.

\begin{figure}[!htbp]
    \centering
    \includegraphics[width=1\textwidth]{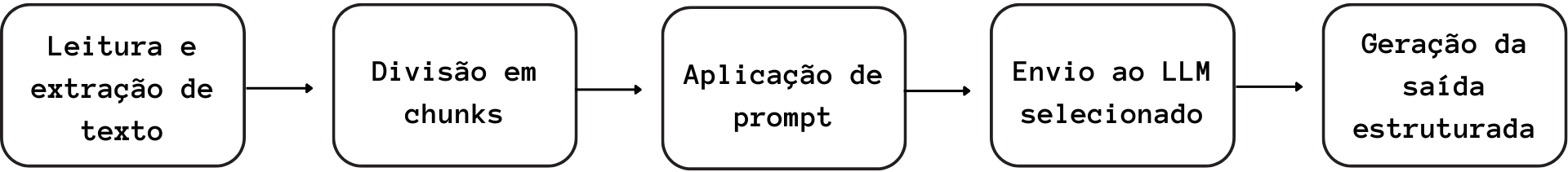}
    \caption{Extraction pipeline}
    \label{fig:fig1.png}
\end{figure}

As illustrated in Figure~\ref{fig:fig1.png}, the process begins with reading the report and extracting textual content while preserving fidelity to the original data. Next, the text is divided into logical blocks (chunks) to maintain the context of each vulnerability within the token limitations imposed by language models.

Each block is processed by a specific prompt that instructs the LLM to identify relevant fields such as description, impact, solution, and references, returning a structured and semantically coherent output. This step accommodates variations across vulnerability scanners and ensures consistency in the extracted results.

Finally, the data undergo validation and consolidation that removes duplicates, checks syntactic conformity, and reconstructs the full vulnerability set. This step ensures information integrity and prepares the final output for integration with anonymization and prioritization modules.

\section{Evaluation}
\label{sec:análise}

For evaluation, we used an OpenVAS technical report containing 34 vulnerabilities of varying severities, from which a manually constructed baseline was validated by two independent analysts. Using this reference set, automated extraction experiments were conducted using GPT-4, GPT-4.1, Llama-3, Llama-4, and DeepSeek, all configured with temperature $T = 0.2$ and average blocks of approximately 9,000 characters, respecting each LLM's token limits. Similarity between extractions and the baseline was assessed using the \textit{ROUGE-L} metric, classifying results into Divergent ($\leq 0.4$), Slightly Similar ($\leq 0.6$), Moderately Similar ($\leq 0.7$), and Highly Similar ($>0.7$).

\begin{figure}[!htbp]
    \centering
    \includegraphics[width=0.95\textwidth]{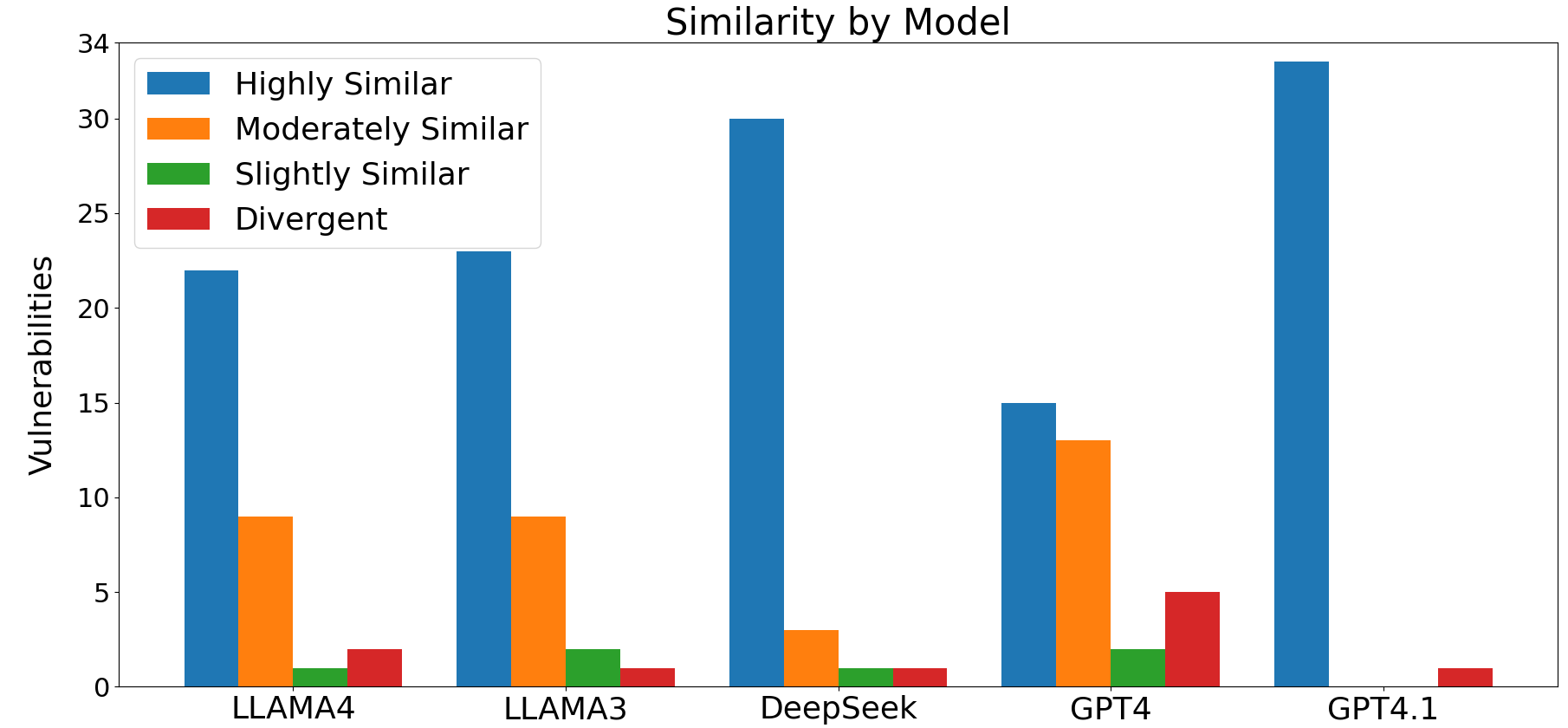}
    \caption{Average similarity (ROUGE-L) between extractions and the baseline}
    \label{fig:fig2.png}
\end{figure}

The results, shown in Figure~\ref{fig:fig2.png}, indicate that DeepSeek and GPT-4.1 achieved the best performance in structured extraction, demonstrating greater capacity for interpretation and preservation of original content. This result can be attributed to more recent architectures, broader training datasets, and optimizations aimed at deep contextual understanding.

In contrast to more robust models, Llama-3 and Llama-4 prioritize computational efficiency and lower operational costs, which may limit performance in complex extraction tasks requiring high semantic consistency. GPT-4 also performed worse than GPT-4.1, reflecting architectural and alignment improvements introduced in the most recent version. Qualitative inspection confirmed occasional inconsistencies such as duplications, omissions, and labeling errors, especially in similar vulnerabilities related to SSL/TLS protocols, as well as substitutions with semantically close terms that harmed exact correspondence with the baseline.

\begin{figure}[!htbp]
    \centering
    \includegraphics[width=0.55\textwidth]{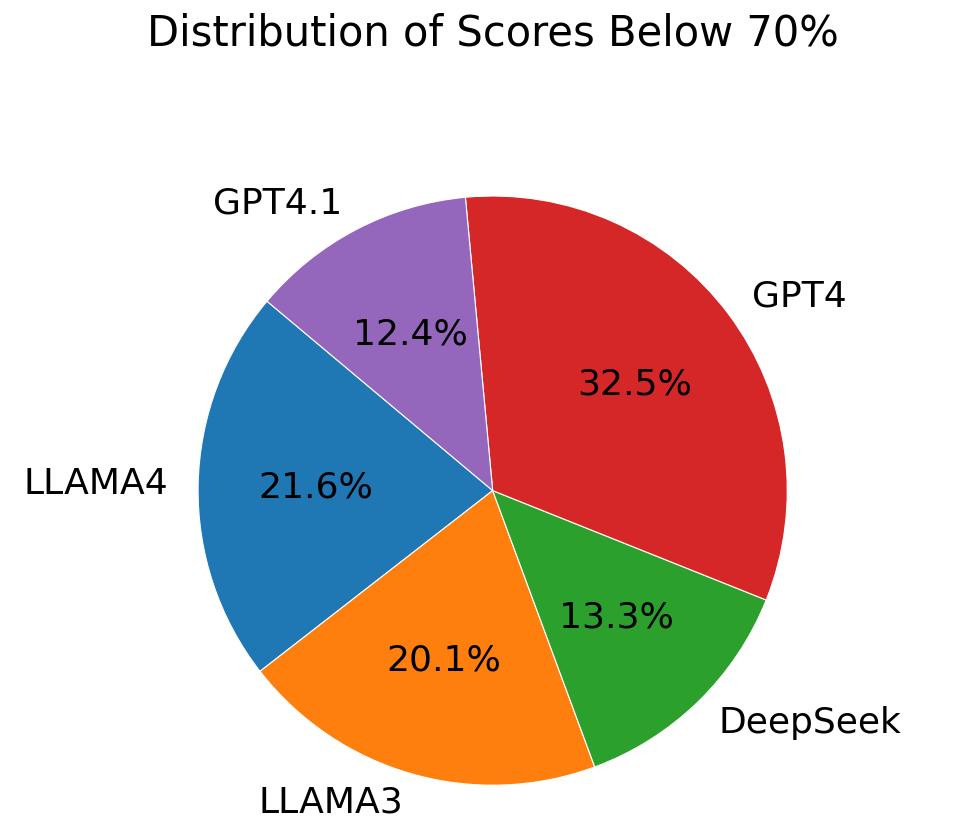}
    \caption{Similarity below Highly Similar}
    \label{fig:fig3.png}
\end{figure}

The analysis of precision degradation, represented by the percentage of fields with similarity below 70\% in Figure~\ref{fig:fig3.png}, identified four main contributing factors: context limitations caused by chunking, which reduce global visibility of the vulnerability; semantic truncation and hallucinations triggered by cuts in technical sections such as \texttt{"NVT:"} or \texttt{"CVSS:"}; and other irregularities such as loss of delimiters and minor tokenization variations between executions. These elements explain the observed discrepancies and highlight the need for more robust segmentation and validation strategies to increase extraction reliability.

Delimiter loss was also identified, recurring during PDF extraction and causing shifts or fragmentations of fields such as \textit{Summary} and \textit{Impact}. Together with random variations arising from tokenization differences or the sampling process, even at lower temperatures, these issues produce inconsistencies between runs. These combined factors explain the observed reduction in precision and reinforce the need for adequate chunking strategies, along with post-processing validation mechanisms, to ensure the integrity and reliability of the extractions.

\section{Final Considerations}
\label{sec:conclusão}

This work presents a method to extract vulnerabilities from OpenVAS and Tenable~WAS reports, producing standardized datasets from unstructured documents by means of prompts with explicit mapping and logical chunking. The results show that models such as GPT-4.1 achieve high similarity relative to the baseline, demonstrating that complex technical reports can be effectively structured to support vulnerability prioritization. As future work, we intend to extend the evaluation to other reports, investigate the use of SLMs, and incorporate mechanisms for anonymizing sensitive data.

\bibliographystyle{IEEEtran}
\bibliography{sbc-template}

\end{document}